\documentclass{article}

    \usepackage[english]{babel}
    \usepackage{authblk} 
    \usepackage[10pt]{moresize} 
    \usepackage{float}
    \usepackage{hhline}
    \usepackage{booktabs}
    \usepackage{tabularx}
    \usepackage{longtable} 
    \usepackage{array}
    \usepackage{hhline}
    \usepackage{multirow}
    \usepackage{blindtext}
    \usepackage{amssymb}
    \usepackage{subcaption} 
    \usepackage[a4paper,top=1.75cm,bottom=1.75cm,left=1.75cm,right=1.75cm,marginparwidth=1.75cm]{geometry}
    \usepackage{amsmath}
    \usepackage{graphicx}
    \usepackage[colorlinks=true, allcolors=blue]{hyperref}

\date{Supercomputing Asia 2024 (SCA24), 19th February, Sydney, Australia}

\title{\textbf{Speed, power and cost implications for GPU acceleration of Computational Fluid Dynamics on HPC systems}}

    \author[1,2,*]{Zachary Cooper-Baldock} 
    \author[1,3,*]{Brenda Vara Almirall}
    \author[3]{Kiao Inthavong}

    \affil[1]{\small National Computational Infrastructure, Australian National University, Canberra, Australia} 
    \affil[2]{\small Centre for Defence Engineering, Research and Training, College of Science and Engineering, Flinders University, Adelaide, Australia}
    \affil[3]{\small Mechanical \& Automotive Engineering, School of Engineering, Royal Melbourne Institute of Technology University, Victoria, Australia}
    \affil[*]{\small These authors had an equal contribution to this body of work} 

\begin{document}
\maketitle


\begin{abstract}
Computational Fluid Dynamics (CFD) is the simulation of fluid flow undertaken with the use of computational hardware. The underlying equations are computationally challenging to solve and necessitate high performance computing (HPC) to resolve in a practical timeframe when a reasonable level of fidelity is required. The simulations are memory intensive, having previously been limited to central processing unit (CPU) solvers, as graphics processing unit (GPU) video random access memory (VRAM) was insufficient. However, with recent developments in GPU design and increases to VRAM, GPU acceleration of CPU solved workflows is now possible. At HPC scale however, many operational details are still unknown. This paper utilizes ANSYS Fluent, a leading commercial code in CFD, to investigate the compute speed, power consumption and service unit (SU) cost considerations for the GPU acceleration of CFD workflows on HPC architectures. To provide a comprehensive analysis, different CPU architectures, and GPUs have been assessed. It is seen that GPU compute speed is faster, however, the initialisation speed, power and cost performance is less clear cut.  Whilst the larger A100 cards perform well with respect to power consumption, this is not observed for the V100 cards. In situations where more than one GPU is required, their adoption may not be beneficial from a power or cost perspective.
\end{abstract}


\section{Introduction}

    Computational fluid dynamics is a subset of traditional fluid mechanics that uses numerical analysis and computational hardware to calculate and resolve fluid flow in 2-dimensional and 3-dimensional space \cite{twodvsthreed}. Computational hardware is used to perform calculations of the freestream flow, fluid interactions and surface effects that are exerted as a fluid medium moves, is heated or otherwise changed. CFD has existed for a relatively short period of time. Douglas Aircraft first developed fluid Panel Methods in 1967, followed by the initial early versions of ANSYS Fluent and STAR-CD’s 3D unstructured CFD codes \cite{futureCFD}. From their inception, the study of fluid has presented computational challenges, as the Navier Stokes equations, the basis of fluid flow, are challenging to resolve. Subsequently CFD at large scale has predominantly run on high performance computing (HPC) systems. This is due to the greater computational resources offered by these facilities, allowing researchers to utilise much higher CPU and RAM allocations than what is typically available at a desktop or workstation level \cite{impactHPC} to accelerate the speed of computation. The basis of these calculations are the Navier-Stokes (N-S) equations \cite{issues}. The N-S equations define the movement of single-phase flow fields involving viscous fluid substances. The direct calculation of the N-S equations directly is computationally intensive \cite{timestep} and is subsequently often simplified or modified \cite{RANSmod} to a particular application or use case, depending on the fluid medium of interest. For compressible flows, such as high-speed applications or compressible gaseous mediums, the Reynolds-Averaged Navier-Stokes (C-RANS) equations are used \cite{compflow}. For incompressible flows, such as water and air, the incompressible Reynolds-Averaged Navier-Stokes (RANS) can be used \cite{RANS}. 
\newline \newline    
    Throughout this calculation process, a CFD analysis has several primary stages. First is the design of geometry and physical boundary conditions. This is followed by the problem domain being decomposed into distinct cells, called the ``mesh''. These mesh cells may be structured or unstructured; or a mixture of the two. The mesh elements may be composed of prismatic, hexagonal or a variety of other shapes, depending on the application. This stage is followed by the development and selection of appropriate flow models, turbulence models and any additional equations, coefficients or numerical values. This stage is often informed by existing literature or prior studies and experimental testing relevant to the specific analysis being undertaken. After this, the simulation is carried out and the equations are solved. The simulation can be time dependent (transient), where the flow is calculated in accordance with a discrete time step, or steady state, where it is iterated irrespective of time until convergence is achieved. Following the successful completion of the simulation, it is post processed and the resulting numerical data is analysed.
\newline \newline
    The most computationally intensive parts of the process are the mesh creation stage and the simulation stage. Traditionally, the simulation stage has taken the longest duration of time. The duration of the simulation extends with increases in the mesh density or increases to the time duration undergoing analysis. The simulation duration can also be extended by the Reynolds number (corresponding to simulation flow speed) \cite{HPCCFD}, however, this increase is typically correlated to greater mesh density requirements at higher speeds \cite{HPCCFD}. Simulations of flow have historically been CPU and RAM driven due to the large size of the computational mesh, which was previously unable to be supported by GPU VRAM \cite{GPULES}. The mesh is required to be stored, either in RAM or VRAM. The sizes of the mesh are model-dependent but can exceed 500M cells in some applications. In recent years, advances the available GPU VRAM has increased to a point where cards such as the NVIDIA A100 have 80GB of HBM2e VRAM \cite{A100}. This is now sufficiently high to store larger mesh cell counts, offering the ability for GPU acceleration to be conducted.
\newline \newline
    This, coupled with the uptake in GPU-enabled HPC hardware has resulted in commercial software, such as ANSYS Fluent, beginning to support GPU accelerated and enabled CFD compute \cite{multiGPU}. Historically, incremental increases in performance and compute speed were seen at greater core counts and RAM sizes \cite{SIEMENS}. This contrasts with the early results of the use of GPUs for CFD, where higher compute speeds were seen, correlating to between 5.23 and 78 times their CPU equivalent according to \cite{multiGPU} and \cite{GPUacc} due to the highly parallel nature of GPU architectures. The provided performance increases, however, are often dimensionless speed increases offered in comparison to a single baseline – which itself is typically a single simulation run on one hardware architecture. This is unhelpful in HPC applications, where there are often a variety of queues and architectures available for use, each with different CPU, RAM and GPU variants. For example, the Australian GADI supercomputer offers 14 different available queues, for job submission to many different architectures \cite{queuestruc}. Without knowing specific quantitative data for a variety of simulations, across a variety of hardware, a specific queue or architecture cannot be targeted.
\newline\newline
    The primary contribution of this paper is a benchmark of GPU acceleration, against traditional architectures, not only for raw speed, as is often reported, but a variety of other important factors. These are the simulation speed, iteration speed, initialisation speed, cost (SU) and power consumption. Two distinctly separate simulations have been used, based on the types of un-optimized models engineers develop for their independent research. The two distinct simulations have been used to provide an understanding of the differences between CFD analysis type (external and internal) as well as the possible impact of turbulence model types, mesh and geometry differences. A broad comparison is required as CFD research encompasses a variety of different use cases, that are often substantially different to the simplified benchmarks provided prior. Speed, power and cost have been assessed, as these factors are important to both researchers and HPC administrators. 


\section{Methodology}

This work intends to make an assessment of several important aspects pertaining to HPC CFD simulation on ANSYS Fluent R1 2023. These are the simulation time, iteration time, initialisation time, power consumption and simulation cost (SU). These are important metrics by which the effectiveness of simulation on a given architecture can be judged. As has been conveyed, CFD simulations are large-scale, intensive endeavours. With compute times of several days, weeks or months, speed is an important metric. The total simulation time and the speed of the iterations within the simulation. A reduction in either, even a small one, can significantly reduce the time spent simulating on an HPC. Initialisation speed is also important. The initialisation of a CFD analysis is a multistep process that can be lengthy. This process covers the mesh import, boundary condition definition, solver setting selection, initialisation, solution initialization, mesh metric checking, and in some applications also a grid adaptation task. Only after these tasks have been performed can the simulation commence. This initialisation process is predominantly a sequential task and does not greatly benefit from a highly parallel architecture. Also of high importance is the power consumption and the cost of conducting the simulation. The power consumption will differ based on the selected architecture, CPU process node, RAM allocation, and GPU type. Faster computational speeds are often seen on high RAM, VRAM or clock speed components, but there is a power consumption implication of running on this hardware. This trade-off will be evaluated during this analysis. The service unit (SU) cost will also be assessed, as this is an important factor when submitting jobs under a predefined budget. Faster simulations at lower SU costs are highly desirable to researchers conducting these studies.

\subsection{Hardware}
    Australia's GADI supercomputer will be used to perform the assessments outlined above. The Australian GADI supercomputer as of January 2024, is a 4,962 node supercomputer comprising of the Intel Sapphire Rapids, Skylake, Broadwell, Cascade Lake and AMD EPYC CPUs, as well as the NVIDIA V100 and DGX A100 GPUs. The system contains more than 250,000 CPU cores, 930 terabytes of memory, and 640 GPUs \cite{GADISPECS}. The tests will be conducted on the Intel Sapphire Rapids, Intel Broadwell, Intel Cascade Lake and AMD EPYC architectures as outlined in Tables \ref{tab:queuestruct}. GPU acceleration, where applicable, will be performed using the NVIDIA Telsa Volta V100 and NVIDIA DGX A100 GPUs. More details on the testing configuration are provided in Table \ref{tab:queuestruct}. 

                                \begin{table} [H] 
                                    \centering
                                    \begin{ssmall}
                                    \begin{tabular}{m{1.2 cm} | m{0.5 cm} | m{0.7cm} | m{3.2 cm} | m{2.0 cm} | m{0.85 cm} | m{2.5cm} | m{0.85cm}}
                                    \toprule \toprule
                                    Identifier  & Core \# & RAM (GB) & CPU & Architecture & TDP (W)& GPU & TGP (W)\\\toprule \toprule
                                    Broadwell    & 14  & 128 & Intel Xeon E5-2690v4     & Broadwell       & 135 (2) & - & - \\\midrule
                                    Sapphire Rapids    & 52  & 256  & Intel Xeon Platinum 847Q & Sapphire Rapids & 350 (2) & - & - \\\midrule
                                    Cascade Lake      & 24  & 96  & Intel Xeon Platinum 8274 & Cascade Lake    & 240     & - & - \\\midrule
                                    V100  & 24  & 64  & Intel Xeon Platinum 8268 & Cascade Lake & 205 & NVIDIA Telsa Volta V100 (SXM2) 32GB & 300 (2) \\\midrule
                                    A100   & 16  & 128 & AMD EPYC 7742            & 7002 series  & 225 & NVIDIA A100 (SXM4) 80GB & 400
                                    \\\bottomrule
                                    \end{tabular}
                                    \end{ssmall}
                                    \caption{Queue structure breakdown of CPU, RAM and GPU allocation}
                                    \label{tab:queuestruct}
                                \end{table}                             
    CFD simulations often take many iterations to achieve full numerical convergence – often in the order of thousands of iterations \cite{iteration}. As this study is wide-ranging, it was decided that the simulations would be conducted with limited iteration counts of 50 - 100 for both the external and internal assessments. This was done to ensure that reasonable compute times were maintained on the slower architectures.

\subsection{Software}

This CFD analysis will be conducted using the ANSYS Fluent commercial package. Specifically, the version used in this analysis is the 2023 distribution, Revision 1 (2023r1). Parallelization will be carried out using the distributed memory model OpenMPI, version 4.1.4. GADI makes use of the Rocky Linux 8 operating system, with storage managed by a DDN Lustre parallel file system.

\subsection{Calculation methods}

\label{calculations}
    \subsubsection{Simulation time}
    The simulation speed will be given by the total time (\textit{$t_{T}$}). This time will be comprised of all factors pertaining to the simulation process. This includes the mesh import time (\textit{$t_{M}$}), boundary condition and solver setting setting time (\textit{$t_{SS}$}), mesh and domain initialisation time (\textit{$t_{init}$}), as well as the sum of the per iteration time (\textit{$t_{iter}$}), for a given number of iterations (\textit{$n_{init}$}). The formula defining the total simulation time is given by Equation \ref{eqn:simtime}.

                                \begin{equation}
                                   t_T = t_M + t_{SS}+t_{init}+\sum_{n=1}^{n_{iter}}t_{iter}
                                    \label{eqn:simtime}
                                \end{equation}
                                
    \subsubsection{Mean iteration time}
    The mean iteration speed (\textit{$t_{M,iter}$})will be given by the sum of all iteration times (\textit{$t_{iter}$}), divided by the number of iterations under analysis (\textit{$n_{iter}$}). The formula defining this is given by Equation \ref{eqn:itertime}.
    
                                \begin{equation}
                                   t_{M,iter} = \frac{\sum_{n=1}^{n_{iter}}t_{iter}}{n_{iter}}
                                    \label{eqn:itertime}
                                \end{equation}
                                
    \subsubsection{Initialisation time}
    The initialisation speed (\textit{$t_{start}$}) is comprised of the steps taken prior to the analysis. This includes the mesh import time (\textit{$t_{M}$}), boundary condition and solver setting setting time (\textit{$t_{SS}$}), mesh and domain initialisation time (\textit{$t_{init}$}). This is given by Equation \ref{eqn:inittime}.

                                \begin{equation}
                                   t_{start} = t_M + t_{SS}+t_{init}
                                    \label{eqn:inittime}
                                \end{equation}
    
    \subsubsection{Power consumption}
    The time-relative power consumption in watt hours (\textit{$P_{Wh}$}) consists of the maximum theoretical power consumption (\textit{$P_{W,max}$}) in watts multiplied by the total simulation time (\textit{$t_{T,h}$}) in hours. It is assumed that the architectures will be at or close to their maximum power consumption. This is assumed due to the good scaling of ANSYS software across multiple cores and the large matrix operation of GPU acceleration. Initial testing on the V100 and A100 cards reported GPU utilisation rates exceeding 85\% according to the NVIDIA-smi command conducted at the node level. Equation \ref{eqn:powercons} provides the relative power consumption and Equation \ref{eqn:powermax} provides the maximum theoretical power consumption. 
                                \begin{equation}
                                   P_{Wh} = P_{W,max}\cdot{t_{T,h}}
                                    \label{eqn:powercons}
                                \end{equation}
    Where:
                                \begin{equation}
                                   P_{W,max} = P_{TDP} \cdot n_{cpus}+P_{TGP} \cdot n_{gpus}
                                    \label{eqn:powermax}
                                \end{equation}
    \subsubsection{Cost}
    Cost is depicted as the service unit cost of using the GADI HPC facilities, as given by Equation \ref{eqn:chargerate}. Like many HPC system providers, this cost consists of the queue charge rate (\textit{$Q_{R,hr}$}), number of CPUs requested (\textit{$n_{cpus}$}) and memory proportion (\textit{$p_{mem}$}) and the walltime used (\textit{$t_{T}$}). The memory proportion is given by Equation \ref{eqn:memport} and consists of the memory requested (\textit{$n_{mem,req}$}) and the memory per core (\textit{$n_{mem,core}$}). The queue charge rate is hardware dependent. The queue charge rate is 1.25 (Broadwell), 2.00 (Sapphire Rapids), 2.00 (Cascade Lake), 3.00 (V100) and 4.50 (A100). 

                                \begin{equation}
                                   C_{SU} = Q_{R,hr} \cdot max(n_{cpus}, p_{mem}) \cdot t_{T}
                                    \label{eqn:chargerate}
                                \end{equation}

                                \begin{equation}
                                   p_{mem} = \frac{n_{mem,req}}{n_{mem,core}} 
                                    \label{eqn:memport}
                                \end{equation}

\subsection{Simulation Design}
To make a practical assessment of speed, power consumption, and cost, two real world simulations have been used. These simulations encompass two common types of analysis: external flows and internal flows. Both simulations use the steady-state RANS equations, at similar cell counts of 11 million cells and 16 million cells for the external and internal case respectively. Further details are provided below. 

\subsubsection{External flow}

\begin{figure}[H]
                                    \centering
                                    \includegraphics[width= 0.80 \textwidth]{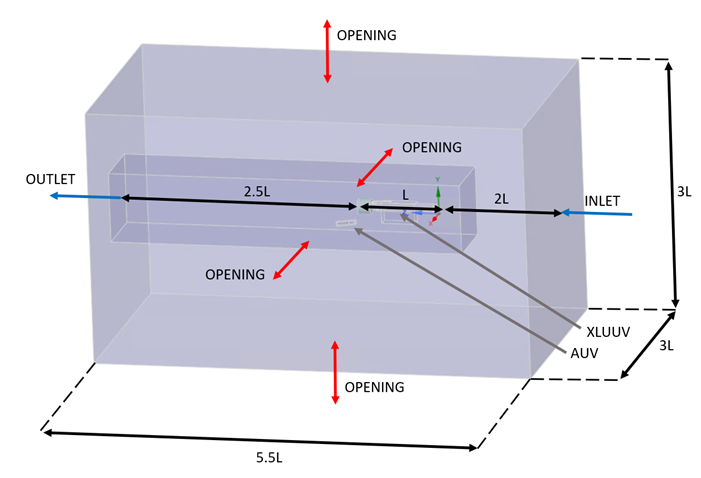}
                                    \caption{3D computational domain. XLUUV indicated in the center of the image.}
                                    \label{fig:xluuvdomain}
                                \end{figure}

External flows are common in many engineering and scientific applications. External flow simulations are undertaken for vessels, vehicles or bodies of interest that are moving through a fluid medium such as air or water. The external flow case for this investigation will be a submarine vessel, modelled on the BOEING Orca platform extra large uncrewed underwater vehicle. An existing mesh convergence study and validation has been provided for this CFD model, at a cell count of 11 million cells \cite{03aad0447576463dafda45c7b7126eea}. The simulation will be a steady state RANS analysis using the realizable k-epsilon model. The geometry of this vessel and the domain is displayed in Figure \ref{fig:xluuvdomain}, with distances defined by the characteristic length of the vessel (\textit{L}) which is 26 meters. Further details on the mesh and geometry can be found in the related publication \cite{03aad0447576463dafda45c7b7126eea}. This simulation takes place at a flow speed of 3 knots (1.55 m/s). A water fluid model has been developed for use in the analysis. This is provided in Table \ref{tab:watermod}. The governing equation of the simulations is the incompressible Reynolds-Averaged Navier-Stokes (RANS) equations.

                                \begin{table} [H] 
                                    \centering
                                    \begin{ssmall}
                                    \begin{tabular}{m{4 cm} | m{3 cm} | m{1.6 cm} }
                                    \toprule \toprule
                                    Fluid parameter                 & Value                             & Reference         
                                    \\\toprule \toprule
                                    Simulation depth                & 100 [m]                           & -                 \\\midrule
                                    Temperature                     & 21.8 [\textdegree C]              & -                 \\\midrule
                                    Pressure                        & 1.0057 [MPa]                      & -                 \\\midrule
                                    Molar mass                      & 18.434 [kg/kmol]                  & \cite{pilson}     \\\midrule
                                    Density                         & 1025.1627 [kg/$m^{3}$]            & \cite{sharqawy}   \\\midrule
                                    Specific heat capacity          & 4003.3 [J/kg.K]                   & \cite{millero}    \\\midrule
                                    Thermal conductivity            & 0.58981 [W/mK]                    & \cite{castelli}   \\\midrule
                                    Dynamic viscosity               & 0.0010314 [kg/m/s]                & \cite{isdale}     \\\midrule
                                    Specific enthalpy               & 86.961 [kJ/kg]                    & \cite{sharqawy}   \\\midrule
                                    Thermal expansion coefficient   & $271.4 \times  10^{6} [K^{-1}]$   & \cite{sverdrup}   \\\bottomrule
                                    \end{tabular}
                                    \end{ssmall}
                                    \caption{Physical parameters of the seawater fluid model for the external flow case}
                                    \label{tab:watermod}
                                \end{table}

\subsubsection{Internal flow case}
\begin{figure}[H]
    \centering
    \includegraphics[width=0.85\linewidth]{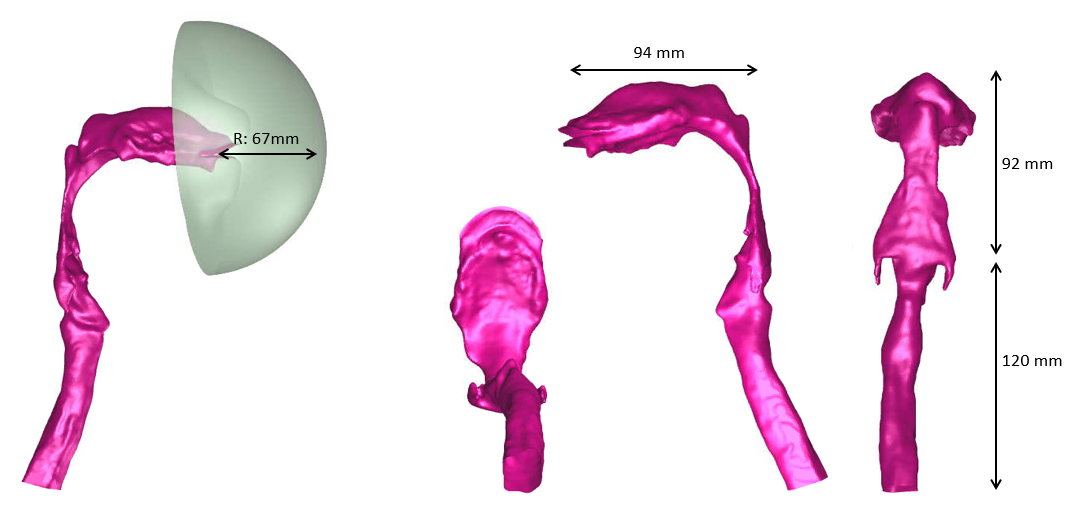}
    \caption{3D computational domain - internal airflow case}
    \label{fig:AJ-dimensions}
\end{figure}

Internal flows are also commonly analysed in engineering applications. The simulations of internal flows are characterised by the bounding object restricting the space available to the fluid. The internal flow case for this investigation was a human mouth-throat airway model, primarily used for assessing drug inhalation. A virtual twin created from a CT scan of a 47-year-old adult male. The geometry of this airway is provided in Figure \ref{fig:AJ-dimensions}. An existing mesh convergence study and validation have been provided for this CFD model, with a cell count of 16 million cells using a poly-hexcore mesh. The simulation was run in steady-state RANS analysis using the k-omega model, and the inhalation boundary conditions replicated an oral inhalation of 30 Liters of air per minute.

\begin{table} [H] 
                                    \centering
                                    \begin{ssmall}
                                    \begin{tabular}{m{4 cm} | m{3 cm} | m{1.6 cm} }
                                    \toprule \toprule
                                    Fluid parameter                 & Value                             & Reference         
                                    \\\toprule \toprule
                                    Space                & 3D                            & -                 \\\midrule
                                    Time                     & Steady                  & -    \\\midrule
                                    Density                         & 1.225 [kg/$m^{3}$] & \cite{tu_computational_2013}            \\\midrule
                                    Pressure                        & 1.0057 [MPa]                
                                        & -   \\\midrule
                                    Dynamic viscosity               & 1.7894e-05 [kg/m/s]                & \cite{tu_computational_2013}      \\\midrule
                                    Specific enthalpy               & 86.961 [kJ/kg]                    & \cite{sharqawy}   \\\midrule
                                    Viscous model   & SST k-omega turbulence model   &  \cite{menter_zonal_1993}   \\\bottomrule
                                    \end{tabular}
                                    \end{ssmall}
                                    \caption{Physical parameters of the air inhaled fluid model for the human airway}
                                    \label{tab:airwaymod}
                                \end{table}


\section{Results}
The results have been provided below for the external and internal simulations. Simulation speed, iteration speed, initialisation speed, power consumption, and relative power to cost are provided. The equations used to determine these values are provided above in Section \ref{calculations}.

\subsection{Simulation time}

\begin{figure}[H]
                                    \centering
                                    \begin{subfigure}{0.48\linewidth}
                                        \centering
                                        \includegraphics[width=\linewidth]{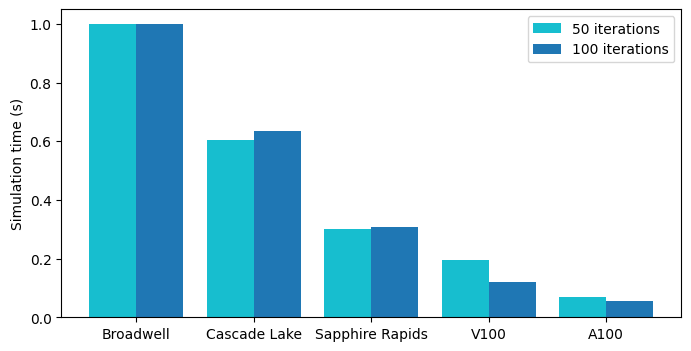}
                                        \caption{External Case}
                                        \label{fig:total_sim_speed_A}
                                    \end{subfigure}
                                    \hspace{0.00\linewidth} 
                                    \begin{subfigure}{0.48\linewidth}
                                        \centering
                                        \includegraphics[width=\linewidth]{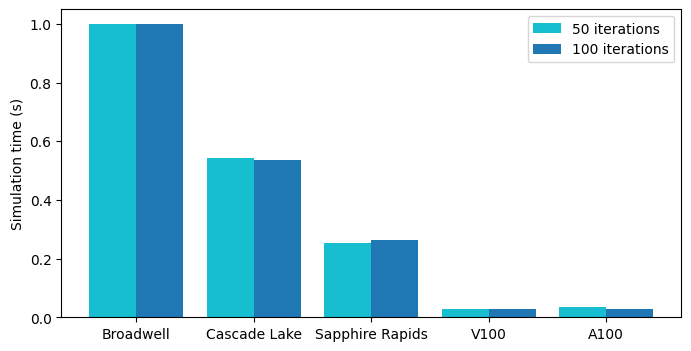}
                                        \caption{Internal Case}
                                        \label{fig:total_sim_speed_B}
                                    \end{subfigure}
                                    \caption{Total simulation time (s) normalised to slowest}
                                    \label{fig:total_sim_speed}
                                \end{figure}

Figure \ref{fig:total_sim_speed_A} and \ref{fig:total_sim_speed_B} detail the total simulation time, normalised to the slowest recorded total time. For the external case (\ref{fig:total_sim_speed_A}), a strong trend in speed is seen. The longest total simulation time is recorded for the Broadwell architecture, at a time of 1192.8 seconds (50 iter.) and a time of 2318.4 seconds (100 iter.). Cascade Lake recorded a total time of 719.2 seconds (50 iter.) and 1468.8 seconds (100 iter.), outperforming Broadwell by 39.7\% (50 iter.) and 42.1\% (100 iter.) respectively. Sapphire Rapids recorded a total simulation time of 358.7 seconds (50 iter.) and 713.9 seconds (100 iter.). GPU acceleration was seen to be faster than all CPU only trials. Notably, there were also discrepancies between the two different GPUs, with the V100 enabled simulation recording total times of 232.2 seconds (50 iter.) and 277.5 seconds (100 iter.) respectively. The larger A100 GPU was 64.4\% faster (50) and 53.7\% faster (100) than the V100, recording times of 82.8 seconds (50 iter.) and 128.5 seconds (100 iter.). 
\newline \newline
For the internal case of the human airway (\ref{fig:total_sim_speed_B}), the same speed trend can be seen. The longest simulation time occured on the Broadwell architecture, taking 3272.1 seconds (50 iter.) and 6946.0 seconds (100 iter.) respectively. The duration of the Cascade Lake was 54\% of Broadwell's taking 1778.3 seconds (50 iter.) and 3716.4 seconds (100 iter.). Sapphire Rapid completed the 50 iterations in 833.8 seconds and the 100 iteration simulation in 1822.8 seconds, the fastest of the CPU only architectures. Resulting in a simulation time that was 26\% of the time it took the Broadwell Architecture. The GPU simulations were notably faster. The V100 architecture was only 2.7\% (50 iter.) at 89.0 seconds and 113.0 seconds, 3.4\% (100 iter.) of the maximum time recorded. The A100 was faster again, completing the simulations in 113.1 seconds with 3\% (50 iter.) and 201.6 seconds or 2.9\% (100 iter.) of the maximum time recorded.

\subsection{Mean iteration time}
\begin{figure}[H]
                                    \centering
                                    \begin{subfigure}{0.48\linewidth}
                                        \centering
                                        \includegraphics[width=\linewidth]{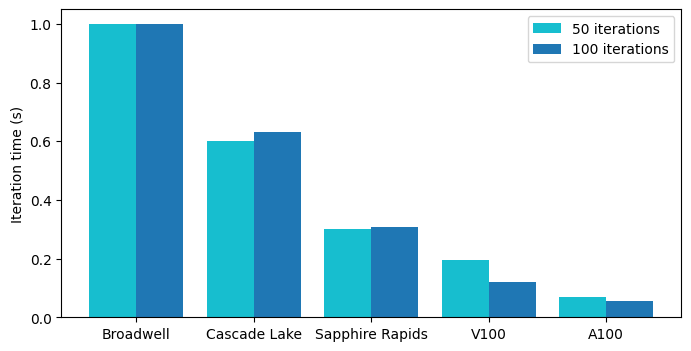}
                                        \caption{External Case}
                                        \label{fig:iter_speed_A}
                                    \end{subfigure}
                                    \hspace{0.00\linewidth} 
                                    \begin{subfigure}{0.48\linewidth}
                                        \centering
                                        \includegraphics[width=\linewidth]{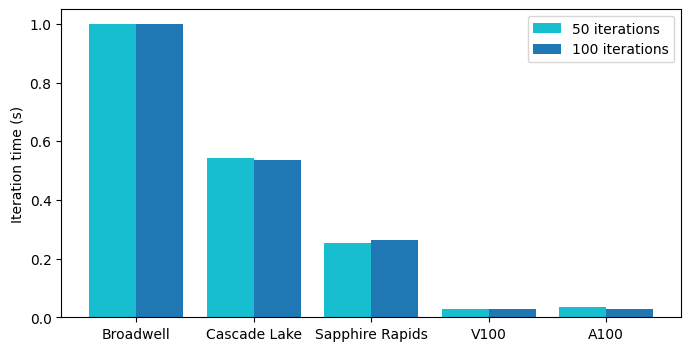} 
                                        \caption{Internal Case}
                                        \label{fig:iter_speed_B}
                                    \end{subfigure}
                                    \caption{Mean iteration time (s) normalised to slowest}
                                    \label{fig:iter_speed}
                                \end{figure}

Figure \ref{fig:iter_speed_A} and \ref{fig:iter_speed_B} detail the individual mean iteration time, normalised to the slowest recorded time. For the external case (Figure \ref{fig:iter_speed_A}), the trend present in total simulation time (Figure \ref{fig:total_sim_speed}) is replicated here. The longest mean iteration time is recorded for the Broadwell architecture, with a time of 23.8 seconds (50 iter.) and a time of 23.1 seconds (100 iter.). Cascade Lake recorded iteration times of 14.3 seconds (50 iter.) and 14.6 seconds (100 iter.) respectively, outperforming Broadwell by 40.0\% (50 iter.) and 37.8\% (100 iter.) respectively. Sapphire Rapids recorded a mean iteration time of 7.2 seconds (50 iter.) and 7.1 seconds (100 iter.). GPU acceleration was seen to be faster than all CPU only trials. Notably, there were also discrepancies between the two different GPUs, with the V100 enabled simulation recording iteration times of 4.6 seconds (50 iter.) and 2.8 seconds (100 iter.) respectively. The larger A100 GPU was 65.3\% faster (50 iter.) and 53.6\% faster (100 iter.) than the V100, recording mean iteration times of 1.6 seconds (50 iter.) and 1.3 seconds (100 iter.) respectively. 
\newline \newline
Again, the internal simulation (\ref{fig:iter_speed_B}) highlights a similar trend to what is seen in the external case (\ref{fig:iter_speed_A}). The Broadwell architecture was the slowest at performing each iteration, in both the 50 iteration and 100 iteration trials. The Broadwell architecture displayed a mean iteration time of 65.4 seconds (50 iter.) and 69.4 seconds (100 iter.). This was followed by the Cascade Lake architecture, complete iterations with a mean time of 35.5 seconds (50 iter.) and 37.1 seconds (100 iter.). The Sapphire Rapid architecture was again the fastest CPU architecture, with a mean iteration time of 16.7 seconds (50 iter.) and 18.2 seconds (100 iter.) respectively. The GPU simulations performed significantly faster. The V100 displayed a mean single iteration time of 1.8 seconds (50 iter.) and 2.3 seconds (100 iter.), while the A100 achieved 2.0 seconds (50 iter.) and 2.0 seconds (100 iter.) respectively.

\subsection{Initialisation time}
\begin{figure}[H]
                                    \centering
                                    \begin{subfigure}{0.48\linewidth}
                                        \centering
                                        \includegraphics[width=\linewidth]{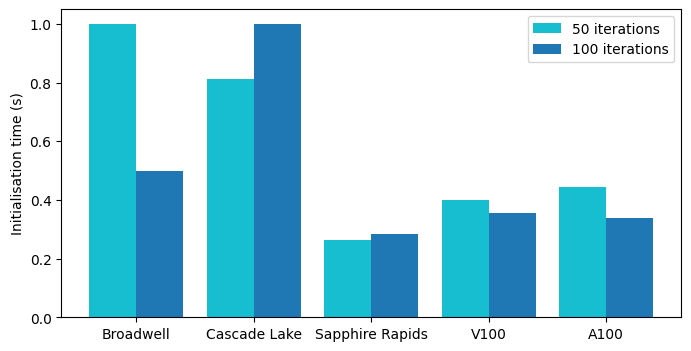}
                                        \caption{External Case}
                                        \label{fig:init_speed_A}
                                    \end{subfigure}
                                    \hspace{0.00\linewidth} 
                                    \begin{subfigure}{0.48\linewidth}
                                        \centering
                                        \includegraphics[width=\linewidth]{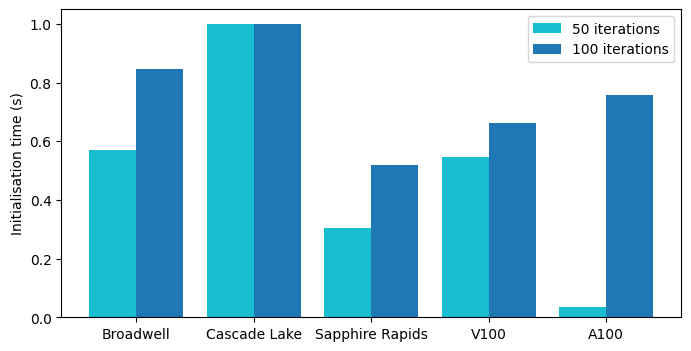} 
                                        \caption{Internal Case}
                                        \label{fig:init_speed_B}
                                    \end{subfigure}
                                    \caption{Initialisation time (s) normalised to slowest}
                                    \label{fig:init_speed}
                                \end{figure}

Figure \ref{fig:init_speed_A} and \ref{fig:init_speed_B} show the initialisation speed. The initialisation speed is the time which is taken to import and structure the mesh file, execute the text commands by the user journal file and begin the simulation. For the external case (\ref{fig:init_speed_A}), a consistent trend was seen for the 50 iteration trials. Initialisation was slowest on the Broadwell architecture at 3.06 seconds, followed by Cascade Lake at 2.49 seconds, then the A100 at 1.36 seconds, V100 at 1.22 seconds. The fastest initialisation time occurred on the Sapphire Rapids architecture at 0.81 seconds. For the 100 iteration trials, the order differs. The slowest initialisation time is recorded for the Cascade Lake architecture at 6.11 seconds, followed by the Broadwell architecture at 3.04 seconds, V100 at 2.18 seconds and A100 with the fastest initialisation speed of 2.07 seconds. 
\newline \newline
For the internal case, the Cascade Lake Architecture had the longest initialisation duration, despite the Broadwell Architecture having a longer total simulation time, as depicted in figure \ref{fig:total_sim_speed_B}.This test saw the fastest initialisation time occur for the A100, at 0.1 seconds. The slowest initialisation time was consistently occurring on the Cascade Lake architecture and was recorded to be 3.48 seconds, for the Cascade Lake 100 iteration trial.

\subsection{Power consumption}
\begin{figure}[H]
                                \centering
                                \begin{subfigure}{0.48\linewidth}
                                    \centering                                    \includegraphics[width=\linewidth]{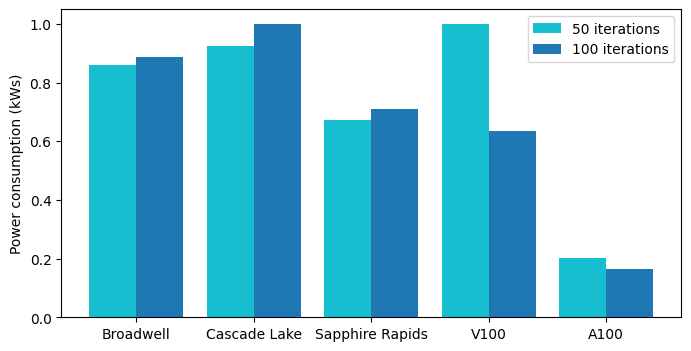}
                                    \caption{External Case}
                                    \label{fig:power_cons_A}
                                \end{subfigure}
                                \hspace{0.00\linewidth} 
                                \begin{subfigure}{0.48\linewidth}
                                    \centering
                                    \includegraphics[width=\linewidth]{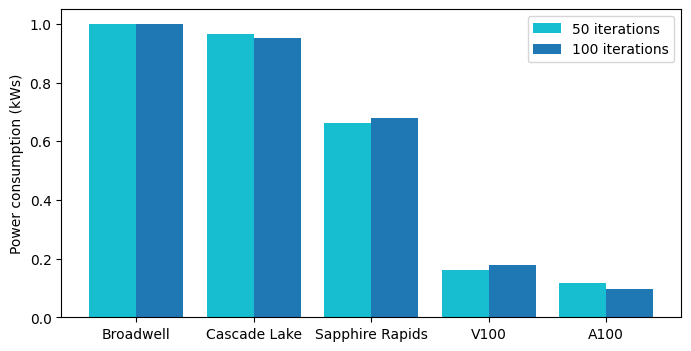} 
                                    \caption{Internal Case}
                                    \label{fig:power_cons_B}
                                \end{subfigure}
                                \caption{Power consumption (Wh) normalised to slowest}
                                \label{fig:power_cons}
                            \end{figure}

Figure \ref{fig:power_cons_A} and \ref{fig:power_cons_B} show the power consumption in watt hours of the architectures under comparison. A full load was assumed for power consumption modelling purposes, where the maximum total wattage of 135W (Broadwell Intel Xeon E5-2690v4), 240W (Cascade Lake Intel Xeon Platinum 8274), 350W (Sapphire Rapids Intel Xeon Platinum 847Q), 805W (Dual V100 cards and Cascade Lake Intel Xeon Platinum 8268) and 456W (A100 and AMD EPYC 7742) were used for each trial respectively. These were then multiplied by the simulation times for 50 and 100 iterations to determine the power consumption. These were then normalised by the highest power consumption. The power consumption results differed between the 50 and 100 iteration tests for the external case (\ref{fig:power_cons_A}). For the 50 iteration tests, the highest power consumption occurred on the V100 architecture at 51.9 Wh, followed by Cascade Lake at 47.9 Wh, Broadwell at 44.7 Wh, Sapphire Rapids at 34.9 Wh and the A100 being the lowest at 10.5 Wh. For the 100 iteration tests, the highest consumption was recorded for the Cascade Lake architecture at 97.9 Wh, Broadwell at 86.9 Wh, Sapphire Rapids at 69.4 Wh, the V100 at 62.0 Wh, with the lowest again recorded on the A100 at 16.3 Wh. The normalised set of these results is provided in Figure \ref{fig:power_cons_A}. 
\newline \newline
The power consumption of the internal case, Figure \ref{fig:power_cons_B}, followed the trends exhibited in the total simulation time (\ref{fig:total_sim_speed_B}) and the mean iteration time (\ref{fig:iter_speed_B}). The slower Broadwell architecture used 122.7 Wh (50 iter.) and 260.0 Wh (100 iter.), while the faster Cascade Lake architecture consumed 118.6 Wh (50 iter.) and 247.8 Wh (100 iter.). The Sapphire Rapid architecture was more energy efficient, due to its lower simulation time, consuming 67\% of the Broadwell architecture power consumption with 81.1 Wh (50 iter.) and 177.2 Wh (100 iter.) respectively. The energy consumption of the GPUs were lower than the CPU architecture, due to their decreased simulation time duration (\ref{fig:total_sim_speed_B}). The relatively higher power draw of the V100, compared to the A100, seen for the external case (\ref{fig:power_cons_A}) was not seen in the internal case (\ref{fig:power_cons_B}). This is likely due to the faster simulation speed of the internal V100 simulations (\ref{fig:total_sim_speed_B}) compared to the external V100 simulations (\ref{fig:total_sim_speed_A}). It was seen that the V100 consumed 19.9 Wh (50 iter.) and 46.3 Wh (100 iter.), whilst the A100 consumed 14.3 Wh (50 iter.) and 25.0 Wh (100 iter.) respectively. The GPUs power draw was between 11.6\% and 17.7\% of the highest that occurred on the Broadwell architecture. 

\subsection{Cost (SU)}
\begin{figure}[H]
                                \centering
                                \begin{subfigure}{0.48\linewidth}
                                    \centering
                                    \includegraphics[width=\linewidth]{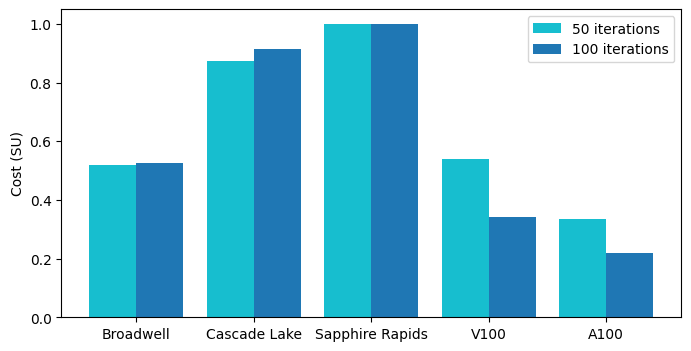}
                                    \caption{External Case}
                                    \label{fig:power_and_cost_A}
                                \end{subfigure}
                                \hspace{0.00\linewidth} 
                                \begin{subfigure}{0.48\linewidth}
                                    \centering
                                    \includegraphics[width=\linewidth]{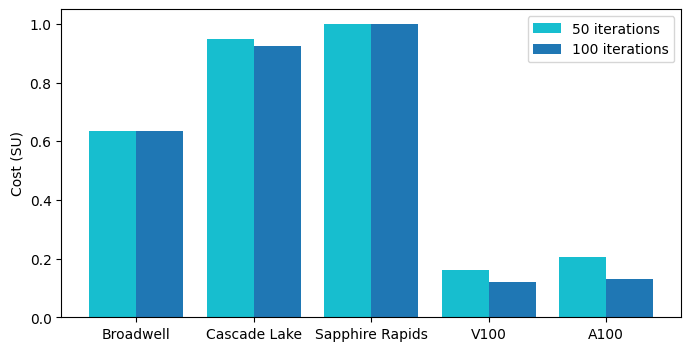} 
                                    \caption{Internal Case}
                                    \label{fig:power_and_cost_B}
                                \end{subfigure}
                                \caption{Service Unit Cost (SU) normalised to slowest}
                                \label{fig:power_and_cost}
                            \end{figure}

Figure \ref{fig:power_and_cost_A} and \ref{fig:power_and_cost_B} denote the service unit (SU) cost associated with the trials conducted for both the external and internal case. For the external case (Figure \ref{fig:power_and_cost_A}), there were again differences in cost between the 50 iteration and 100 iteration trials. For the 50 iteration tests, the most expensive job to run was the Sapphire Rapids (12.8 SU), followed by the Cascade Lake (11.2 SU), V100 (6.9 SU), Broadwell (6.7) and the cheapest was the A100 at 4.3 SU. The V100 was more cost efficient when the number of iterations was increased. At 100 iterations, the most expensive was the Sapphire Rapids (23.1 SU), then Cascade Lake (21.1 SU), Broadwell (12.2 SU), V100 (7.9 SU) and again the cheapest was the 5.1 SU A100 tests. The normalised set of external simulation costs are provided in Figure \ref{fig:power_and_cost_A}. 
\newline \newline
The internal case cost trend (\ref{fig:power_and_cost_B}) replicated that of the external case (\ref{fig:power_and_cost_A}). The queue charge rates had a significant effect, where the time saving nature of the Sapphire Rapid architecture, is not sufficient to negate the SU cost, despite the shorter duration of the simulation. The Sapphire Rapid simulation cost 26.5 SU (50 iter.) and 55.2 SU (100 iter.) and the Cascade Lake architecture incurred costs of 25.2 SU (50 iter.) and 51.0 SU (100 iter.) respectively. The Broadwell architecture was 36\% cheaper at 16.9 SU (50 iter.) and 35.0 SU (100 iter.). The V100 cost was 4.3 SU (50 iter.) and 6.6 SU (100 iter.) respectively, with the cost to use the A100 marginally higher at 5.4 SU (50 iter.) and 7.1 SU (100 iter.). Even with the comparatively high charge rates of the GPU queues, the short total simulation durations ensured it was significantly cheaper to run the GPU simulations.

\section{Discussion}
The acceleration of CFD simulations has been the primary focal point of GPU use in fluid dynamics. The trends in acceleration have been replicated here in Figure \ref{fig:total_sim_speed} \& \ref{fig:iter_speed}, where the increase in compute speed is clear. As expected, with respect to computational speed, the acceleration of the simulations yielded impressive reductions in time taken. Both the V100 and A100 cards resulted in significantly faster simulations, reducing the time taken by 80\% or greater for the external case and 95\% or greater for the internal case. This, when coupled with a similar trend in iteration time, further cements GPU acceleration as the path forward. However, this assessment has also sought to determine the initialisation time, power consumption and cost implications of GPU acceleration. These have found to be less clear cut than that of raw speed. In addition, prior to the undertaking of this investigation, there were discrepancies identified within the existing ANSYS Fluent GPU acceleration literature, where HPC job submission scripts were using incorrect, or ineffective submission flags. The use of these incorrect flags would result in limited GPU memory use, leading to diminished benefits of GPU acceleration. To correctly replicate the results seen in this work, the flag “-gpuapp” should be used when calling ANSYS Fluent the Unix shell files. To aid researchers in future endeavours, Appendix \ref{supplementarydata} provides correct CPU and GPU job submission scripts.
\newline \newline
Initialisation speed is of interest to computational fluid dynamics researchers using commercial code like ANSYS Fluent. This is because there is limited access to the underlying code, and as such, this stage of the code execution cannot be easily changed, optimised or accelerated. For simulations requiring large computational meshes, large lists of commands or complicated initialisation and setup of domain conditions, this portion of the total simulation time can become significant. The results presented (Figure \ref{fig:init_speed}) indicate that the Cascade Lake architecture is notably slower, both in the external and internal case. Users undertaking simulations with large computational meshes, large lists of commands or complicated initialisation and setup of domain conditions, would be well advised to make use of the Sapphire Rapids, V100 or A100 architectures. 
\newline \newline
For the administrators of HPC facilities the power consumption of these simulations is important. Increased power consumption directly correlates to higher energy requirements, operational costs and facility cooling requirements. For the external cases, shown in Figure \ref{fig:power_cons_A}, saw the 50-iteration test produce unexpected results in relation to the V100 simulation, where it recorded the highest Wh power consumption. Due to the significant simulation time reduction shown in Figure \ref{fig:total_sim_speed_A}, it was expected that this would correspond to a lower energy consumption. However, the increased duration of the 50-iteration simulation compared to the A100, and coupled with the requirement of dual V100 cards, to have sufficient VRAM, explained the increased power consumption shown here. This was not reflected in the internal cases (Figure \ref{fig:power_cons_B}) which had more consistent GPU acceleration, leading to a shorter simulation duration and subsequent power consumption. As expected, CPU trials, due to their longer duration, even when coupled with lower CPU only power consumption, are not as energy efficient as GPUs in most cases. Users should carefully assess both VRAM requirements, simulation mesh sizes and available GPU configurations and architectures when aiming for energy efficient simulations.
\newline \newline
Cost is of key importance to users operating within their grant, institutional or private funding arrangements. The service cost is comprised of architecture (queue charge rate), CPU core and memory requirements, and the simulation duration. Users wishing to minimise their service cost would be well suited to follow the trends seen in Figure \ref{fig:power_and_cost}. The CPU only architectures are of consistently higher cost than that of the GPU simulations, across both the external and internal cases, for both 50 and 100 iterations. The Broadwell, Cascade Lake and Sapphire Rapids queues are the more expensive than the V100 and A100 queues. For the external case, the Broadwell queue is relatively close in cost to the V100 and A100 queues, but this is likely due to the longer duration of the V100 external simulations, compared to the internal case, resulting in a higher cost. Users would still be well advised in selecting the V100 or A100 architectures for external simulations. 
\newline \newline
For the internal case (Figure \ref{fig:power_and_cost_B}), as the V100 simulations were of similar duration to the A100 simulations (Figure \ref{fig:total_sim_speed_B}) unlike in the external case, a lower V100 cost was seen. However, as seen in the external case, the internal case Sapphire Rapids simulations incurred the the highest costs. Indicating that even whilst the Sapphire Rapids offer the fastest CPU only simulation speeds (Figure \ref{fig:total_sim_speed}), the queue charge rate and simulation duration results in higher costs. 
\newline \newline
This body of work highlights the importance of careful selection of architecture and hardware, depending on a researchers’ specific simulation requirements. Inappropriate selection of hardware, architecture or resources will correlate to inefficient simulations – resulting in slower, less efficient, more costly simulations. Whilst GPU acceleration remains a clearly optimal choice, the selection of individual GPUs is less so. Users with smaller mesh requirements may see practical benefits such as reduced cost from single GPU use, however larger models benefit significantly from the adoption of larger but more costly, GPU architectures. 


\section{Conclusions}

In this work, two separate computational fluid dynamics simulations were developed using the commercial CFD software ANSYS Fluent. These two simulations were developed to provide a practical assessment of speed, power and cost implications of utilising GPU acceleration on a HPC system. The first simulation was an external fluid flow simulation, developed using a submarine geometry, whilst the second is an internal fluid flow simulation, developed using a model of the human mouth-throat airway. The two simulations were used to assess a single CPU trial on the Broadwell, Cascade Lake and Sapphire Rapids architectures, and to then compare these CPU only simulations to ones accelerated with NVIDIA’s V100 and A100 GPU cards. The simulations were monitored for their total simulation duration, calculation iteration speed, problem and domain initialisation speed, power consumption and cost. 
\newline \newline
The primary findings of this work are multifaceted and are relevant to a variety of different audiences. The study highlights that initialization time is a crucial factor for researchers using commercial codes like ANSYS Fluent, where limited access to underlying code makes optimization challenging. The Broadwell and Cascade Lake architectures, as demonstrated, are notably slower in this regard, prompting researchers initialising complex simulations to opt for architectures such as the Sapphire Rapids, V100, or A100. For researchers constrained by grant or institutional funding pools, cost reduction is also significant motivation. Despite the higher queue charge rates, the speed of GPU acceleration proved more cost-effective than CPU-only approaches, with the V100 and A100 architectures standing out as favourable choices. If the researchers seek the lowest CPU-only cost, the Broadwell architecture was notably more cost effective than either the Cascade Lake or Sapphire Rapids architecture. 
\newline \newline
Power consumption, whilst primarily the concern of HPC facility management, brings important findings in relation to specific GPU architectures. Although more energy efficient than the CPU architectures, the V100 recorded high power consumption in certain scenarios. With this underscoring the importance of careful assessment of VRAM requirements, simulation mesh sizes, in order to select appropriate GPU configurations for energy-efficient HPC facilities and long duration computational fluid dynamics simulations. While GPU acceleration presents a promising speed, power, and cost trajectory for CFD simulations run on ANSYS Fluent. A comprehensive understanding of one’s specific simulation is required to effectively and efficiently conduct work on HPC facilities. These results hope to provide a starting point for researchers looking to conduct ANSYS Fluent CFD simulations on HPC facilities with or without GPU acceleration as well as other CFD software. Future research should seek to quantify these results for more intensive simulations, such as those with particulate-laden flow, to see if similar trends emerge.  


\section{Acknowledgements}
This research was undertaken with the assistance of resources and services from the National Computational Infrastructure (NCI), which is supported by the Australian Government.
\newline \newline
This work was completed in part at the NCI Open Hackathon June 2023, part of the Open Hackathons program. The authors would like to acknowledge OpenACC-Standard.org for their support. 
\newline \newline
We also could not have completed this work without Fredrick Fung, who connected the two primary authors of this paper. Acting as a catalyst in the creation of this work. 

\bibliographystyle{ieeetr}
\bibliography{sample}

\begin{thebibliography}{10}

\bibitem{twodvsthreed}
M.~G. Perez and E.~Vakkilainen, ``A comparison of turbulence models and two and
  three dimensional meshes for unsteady cfd ash deposition tools,'' {\em Fuel},
  vol.~237, no.~1, pp.~806--811, 2018.

\bibitem{futureCFD}
{Amazon Web Services}, ``The future of \uppercase{CFD},'' 2023.

\bibitem{impactHPC}
C.~Lange, P.~Barthelmas, T.~Rosnitchek, S.~Tremmel, and F.~Rieg, ``Impact of
  \uppercase{HPC} and automated \uppercase{CFD} simulation processes on virtual
  product development — a case study,'' {\em Applied Sciences}, vol.~11,
  no.~14, p.~6552, 2021.

\bibitem{issues}
P.~M. Gresho, ``Some current cfd issues relevant to the incompressible
  navier-stokes equations,'' {\em Computer Methods in Applied Mechanics and
  Engineering}, vol.~87, no.~2, pp.~201--252, 1991.

\bibitem{timestep}
J.~L. Guermond and P.~D. Minev, ``High-order time stepping for the
  navier–stokes equations with minimal computational complexity,'' {\em
  Journal of Computational and Applied Mathematics}, vol.~310, no.~1,
  pp.~92--103, 2017.

\bibitem{RANSmod}
P.~Spalart, ``Reflections on rans modelling,'' {\em Progress in Hybrid RANS-LES
  Modelling}, vol.~1, no.~1, pp.~7--24, 2010.

\bibitem{compflow}
G.~Choubey and M.~Tiwari, ``Pedagogy for the computational approach in
  simulating supersonic flows,'' {\em Scramjet Combustion: Fundamentals and
  Advances}, vol.~1, no.~1, pp.~163--181, 6.

\bibitem{RANS}
B.~Sun, ``Revisiting the reynolds-averaged navier–stokes equations,'' {\em
  Open Physics}, vol.~19, no.~1, pp.~853--862, 2021.

\bibitem{HPCCFD}
S.~J. Lawson, M.~Woodgate, R.~Steiji, and G.~N. Barakos, ``High performance
  computing for challenging problems in computational fluid dynamics,'' {\em
  Progress in Aerospace Sciences}, vol.~52, no.~1, pp.~19--29, 2012.

\bibitem{GPULES}
B.~Papp, G.~Kristof, and C.~Gromke, ``Application and assessment of a gpu-based
  les method for predicting dynamic wind loads on buildings,'' {\em Journal of
  Wind Engineering and Industrial Aerodynamics}, vol.~217, no.~1, p.~104739,
  2021.

\bibitem{A100}
{NVIDIA}, ``\uppercase{NVIDIA} \uppercase{A100} tensor core \uppercase{GPU},''
  2023.

\bibitem{multiGPU}
{G. Petrone}, ``Unleashing the power of multiple gpus for \uppercase{CFD}
  simulations,'' 2022.

\bibitem{SIEMENS}
{S. Gross}, ``An engineer’s guide to the \uppercase{CFD} hardware galaxy,''
  2023.

\bibitem{GPUacc}
Y.~Xiang, B.~Yu, Q.~Yuan, and D.~Sun, ``\uppercase{GPU} acceleration of
  \uppercase{CFD} algorithm: \uppercase{HSMAC} and \uppercase{SIMPLE},'' {\em
  International Conference on Computational Science}, vol.~ICCS 2017, no.~12-14
  June 2017,, pp.~Zurich, Switzerland, 2017.

\bibitem{queuestruc}
{Y. Sun}, ``Queue structure,'' 2023.

\bibitem{GADISPECS}
A.~N. University, ``Hpc systems,'' 2023.

\bibitem{iteration}
S.~Koziel and L.~Leifsson, ``Multi-level \uppercase{CFD}-based airfoil shape
  optimization with automated low-fidelity model selection,'' {\em 2013
  International Conference on Computer Science}, vol.~18, no.~1, pp.~889--898,
  2013.

\bibitem{03aad0447576463dafda45c7b7126eea}
Z.~Cooper-Baldock, P.~Santos, R.~Brinkworth, and K.~Sammut, ``Payload bay
  berthing of underwater vehicles with a larger xluuv,'' in {\em 7th Submarine
  Science Technology and Engineering Conference 2023 - Proceedings SubSTEC7},
  pp.~97--104, Submarine Institute of Australia, Sept. 2023.

\bibitem{pilson}
M.~E.~Q. Pilson, ``An introduction to the chemistry of the sea,'' {\em
  Cambridge University Press}, pp.~--, 2013.

\bibitem{sharqawy}
M.~H. Sharqawy, J.~H.~V. Lienhard, and S.~M. Zubair, ``Thermophysical
  properties of seawater: a review of existing correlations and data,'' {\em
  Desalination and Water Treatment}, vol.~16, no.~1, pp.~354--380, 2010.

\bibitem{millero}
F.~J. Millero, G.~Perron, and J.~E. Desnoyers, ``Heat capacity of seawater
  solutions from 5° to 35°c and 0.5 to 22\% chlorinity,'' {\em Journal of
  Geophysics Research}, vol.~78, no.~21, pp.~4499--4507, 1973.

\bibitem{castelli}
V.~J. Castelli, E.~M. Stanley, and E.~C. Fischer, ``The thermal conductivity of
  seawater as a function of pressure and temperature,'' {\em Deep Sea Research
  and Oceanographic Abstracts}, vol.~21, no.~4, pp.~311--319, 1974.

\bibitem{isdale}
J.~D. Isdale, ``Viscosity of simple liquids including measurement and
  prediction at elevated pressure,'' {\em Doctor of Philosophy, The University
  of Strathclyde}, pp.~--, 1976.

\bibitem{sverdrup}
H.~U. Sverdrup, M.~W. Johnson, and R.~H. Fleming, ``The oceans, their physics,
  chemistry and general biology,'' {\em New York, Prentice-Hall}, pp.~--, 1942.

\bibitem{tu_computational_2013}
J.~Tu, K.~Inthavong, and G.~Ahmadi, {\em Computational {Fluid} and {Particle}
  {Dynamics} in the {Human} {Respiratory} {System}}.
\newblock Biological and {Medical} {Physics}, {Biomedical} {Engineering},
  Dordrecht: Springer Netherlands, 2013.

\bibitem{menter_zonal_1993}
F.~R. Menter, ``Zonal two {Equation} k-w turbulence models for aerodynamic
  flows,'' {\em AIAA Journal}, vol.~32, pp.~1598--1605, July 1993.

\end{thebibliography}
\section{Appendix}
\label{supplementarydata}
Supplementary data is provided at the following GitHub repository. This repository includes the shell files used to conduct the CPU and GPU trials, as well as the correct flag to ensure effective use of the GPU architecture on the fluent call. The repository is located here: \url{https://github.com/Zachary-Cooper-Baldock/SCA24_GPU_Acceleration_of_CFD.git}.

\end{document}